\documentclass[aps,prl,twocolumn,superscriptaddress]{revtex4-2}
\usepackage{amsmath,amssymb,amsfonts}
\usepackage{bm, latexsym}
\usepackage{graphicx}
\usepackage{bbm,times}
\usepackage[normalem]{ulem}
\usepackage[usenames,dvipsnames]{color}
\usepackage[pdftex]{hyperref}
\usepackage{braket}
\usepackage{mathtools}

\begin{document}

\newcommand{\cred}[1]{\textcolor{red}{#1}}

\title{Exact Dynamic Scaling of Quantum Jump Statistics in Monitored Many-Body Systems}



\author{Kazuki Yamamoto}
\thanks{These two authors contributed equally to this work.}
\email{kazuki-yamamoto@omu.ac.jp}
\affiliation{Research Institute for Innovation and Co-Creation, Osaka Metropolitan University, Sakai, Osaka 599-8531, Japan}
\affiliation{Department of Physics, Osaka Metropolitan University, Sumiyoshi, Osaka 558-8585, Japan}
\affiliation{Nambu Yoichiro Institute of Theoretical and Experimental Physics (NITEP), Osaka Metropolitan University, Sumiyoshi, Osaka 558-8585, Japan}
\author{Taiki Ishiyama}
\thanks{These two authors contributed equally to this work.}
\email{ishiyama.t.d861@m.isct.ac.jp}
\affiliation{Department of Physics, Institute of Science Tokyo, Meguro, Tokyo 152-8551, Japan}
\author{Toranosuke Matsubara}
\affiliation{Department of Physics, Institute of Science Tokyo, Meguro, Tokyo 152-8551, Japan}
\author{Kazuya Fujimoto}
\affiliation{Department of Physics, Institute of Science Tokyo, Meguro, Tokyo 152-8551, Japan}
\affiliation{Institute for Liberal Arts, Institute of Science Tokyo, Ichikawa, 272-0827, Japan}
\author{Tomohiro Sasamoto}
\affiliation{Department of Physics, Institute of Science Tokyo, Meguro, Tokyo 152-8551, Japan}
\author{Ryusuke Hamazaki}
\affiliation{Nonequilibrium Quantum Statistical Mechanics RIKEN Hakubi Research Team, RIKEN Pioneering Research Institute (PRI), RIKEN iTHEMS, Wako, Saitama 351-0198, Japan}



\date{\today}

\begin{abstract}
Quantum measurement has opened a new frontier in nonequilibrium many-body physics, revealing striking phenomena such as measurement-induced entanglement transitions, whose experimental observation typically requires costly postselection.
Quantum jump statistics offer another route to studying such measurement-induced many-body phenomena without postselection, motivating the search for universality in measurement outcomes themselves.
In this Letter, we discover that fluctuation of the number of quantum jumps within a subsystem exhibits novel dynamic scaling, characterized by diffusive time scale and anomalous super-Poissonian system-size scaling that depend on the subsystem size in monitored many-body systems.
For free fermions under continuous monitoring, we exactly determine the scaling exponent and the scaling function by deriving an exact finite-size solution of the Liouvillian autocorrelation function.
We also analytically show that the dynamic scaling with the same scaling exponent and the scaling function even holds for interacting fermions in the limit of large measurement strength. 
Our work provides a novel framework to classify measurement-induced nonequilibrium many-body universality through experimentally friendly, postselection-free quantities.
\end{abstract}

\maketitle

\textit{Introduction}--The role of measurement in open quantum systems has attracted considerable attention, as readout of measurement outcomes gives rise to unique nonequilibrium physics beyond conventional Lindblad dynamics~\cite{Daley14, Ashida20, Harrington22, Fisher23rev, Hamazaki26rev}.
In particular, the statistics of quantum jumps under measurement \cite{Raphael15, Garrahan18rev, Landi24, Garrahan10, Carollo19, Causer25, Brandes08, Brandes16, Chia17, Landi21, Landi23, Coppola24, Radaelli26, Santos25, Sifft25} encode trajectory-resolved information that is washed out by ensemble averaging, thereby offering a wide range of research arena, from Glauber coherence in photo-detection \cite{Glauber63} 
to thermodynamic uncertainty relations \cite{Barato15, Garrahan17, Horowitz20, Hasegawa20}.
Recently, the impact of many-body effects on quantum jump statistics has been actively investigated, revealing a rich variety of phenomena~\cite{Znidaric14, Hickey13, Lesanovski13, Carollo18A, Bao24, Cenap12, Buca14, Znidaric14B, Znidaric14E, Landi22, Matsumoto25, Rajabi13, Ptaszynski17, Klein21, Fresco24, Le24, Soares25, Yamamoto26, Salatino26, Fitzner26} such as dynamical heterogeneity in space-time fluctuations of ancillary measurements \cite{Cech24, Cech25, Cea26, Cech26, Cech27} and anomalous system-size scaling of subsystem fluctuations in continuously monitored dynamics \cite{Yamamoto25}.
Crucially, in contrast to paradigmatic measurement-induced many-body phenomena such as entanglement transitions \cite{Fisher18, Smith19, Skinner19, Yaodong19, Noel22, Koh22, Google23, Tang20, Fuji20, Szyniszewski20, Minato22, Muller22, Buchhold21, Marcin23, Mochizuki25, Matsubara25, Matsubara26, Fava23, Oshima25, Zhenyu26, Lavasani21, Kells23}, jump statistics can be accessed solely from the spatiotemporal coordinates of quantum jumps, $\{t_i, x_i\}$, and therefore do not suffer from the exponential overhead associated with postselection \cite{Gullans20L, Yaodong23, Altman24, Ippoliti21, Grover21, Yamamoto23, Mog23, Passarelli24, Max24, Barratt22L1, Agrawal24, Ippoliti24}.

On another front, identifying universal properties in nonequilibrium systems is essential for achieving a unified understanding of many-body phenomena, and the classification of universality classes has emerged as a central problem both in classical \cite{Cardy96, Hohenberg77, Bray94, Chuang91, Nagaya92, Seul94, Deutschlander15, Takeuchi10, Takeuchi11, Takeuchi12} and quantum systems \cite{Lamporesi13, Navon15, Braun15, Clark16, Nicklas15, Eigen18, Takeuchi25, Fujimoto26, Ishiyama25fluctuation}.
Dynamic scaling $F(L,t)=L^\alpha f(t/L^z)$ plays a paradigmatic role in such a classification, where $L$ is the system size, $t$ is the time, $f$ is the scaling function, and $\alpha$ and $z$ are universal exponents.
For example, the growth of surface roughness \cite{KPZ86, Amar92, Takeuchi18} shows the dynamic scaling whose exponents characterize universality classes such as the Kardar-Parisi-Zhang and Edwards-Wilkinson classes, which have recently been discussed even for quantum many-body systems~\cite{Fujimoto20, Fujimoto21, Fujimoto22, Kwon26}.
Recent cold-atom experiments have established dynamic scaling as an experimentally accessible signature of nonequilibrium quantum many-body universality, ranging from phase ordering \cite{Erne18, Prufer18, Glidden21, Huh24, Gazo25} to turbulent cascades~\cite{Galka22}.

More broadly, the quest to characterize nonequilibrium universality has also extended to open quantum systems~\cite{Sieberer25} and, more recently, to monitored many-body systems, e.g., through conformal field theory for measurement-induced criticality~\cite{Jian20,Li21,Zabalo22,Kumar24}.
Despite such progress, it remains unclear how the universality of quantum jump statistics, a fundamental and experimentally friendly probe of monitored dynamics, should be classified in monitored many-body dynamics. 
Recently, Ref.~\cite{Yamamoto25} has conjectured nontrivial and robust system-size scaling of the fluctuations; however, the study was limited to long-time saturation regime and based on small-size numerical simulations. 
This motivates asking whether one can analytically verify more general structures of universality of quantum jump statistics, especially dynamic scaling, which provides a distinct route to understand nonequilibrium universality in monitored quantum matter.

\begin{figure}[tb]
\includegraphics[width=8.5cm]{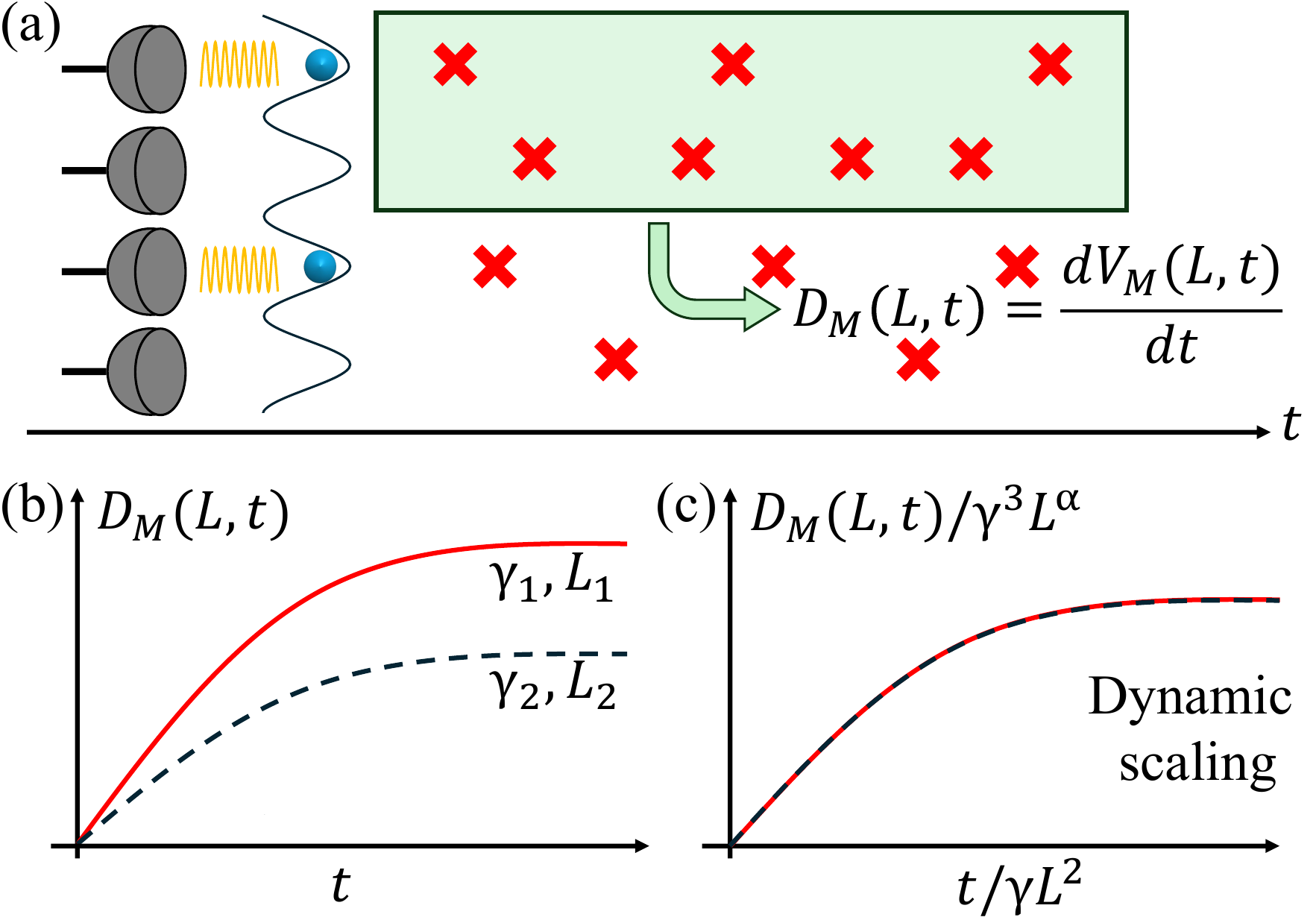}
\caption{(a) Schematic illustration of our setup. We count the number of jumps (red crosses) in a subsystem of size $M$ and obtain its variance $V_M(L,t)$ and noise $D_M(L,t)\equiv \frac{dV_M(L,t)}{dt}$ along trajectory realizations. (b) Time
evolution of the noise for two different parameter sets, $(\gamma_1,L_1)$ and $(\gamma_2,L_2)$. The noise grows with time and eventually saturates. (c) Dynamic scaling of the noise. When we normalize the vertical and horizontal axes by $\gamma^3 L^\alpha$ and $\gamma L^2$, respectively, all curves exhibit a scaling collapse.}
\label{fig_schematic}
\end{figure}

In this Letter, we analytically show for the first time that dynamic scaling characterized by universal exponents emerges in the subsystem noise and variance of the number of quantum jumps [see red crosses in Fig.~\ref{fig_schematic}(a)], where the noise is defined by the time derivative of the variance. 
Specifically, for a free-fermion chain under continuous monitoring, we demonstrate that the subsystem noise $D_{M}$ and variance $V_M$ exhibit super-Poissonian statistics obeying the dynamic scaling $D_{M}(L,t)\simeq\gamma^3 L^\alpha f_M(t/(\gamma L^2))$ and $V_{M}(L,t)\simeq \gamma^3 L^\alpha t F_M(t/(\gamma L^2))$, where $\gamma$ is the measurement strength and $M$ is the subsystem size [see Eqs.~\eqref{eq:asym_loc} and \eqref{eq:asym_half}].
The scaling exponent $\alpha$ and the scaling functions $f_M$ and $F_M$ are determined for arbitrary measurement strengths, demonstrated by an exact finite-size solution of the Liouvillian autocorrelation function.
Furthermore, for sufficiently strong measurement strength $\gamma \gg 1$, we elucidate that the dynamic scaling with the same scaling exponent and the scaling function even holds for continuously monitored interacting fermions.
Because our results can be extracted solely from the spacetime records of quantum jumps, $\{t_i,x_i\}$, our approach requires no postselection and is therefore directly relevant to experiments, as numerically tested with the use of quantum trajectory methods.
Our finding of measurement-induced dynamic scaling establishes a new framework for classification of universality of monitored quantum many-body dynamics through jump statistics.
See Fig.~\ref{fig_schematic} for the summary of our results.

\textit{Setup}--We consider a system of $L/2$ fermions on a one-dimensional chain of length $L\in 2\mathbb N$ under continuous monitoring of the local particle number. The system Hamiltonian is given by $H=-J\sum_{j=1}^{L}(c_{j+1}^{\dagger}c_j+c_j^{\dagger}c_{j+1})$, where $c_j$ is the fermionic annihilation operator, and $c_{L+1}=c_1$ with periodic boundary condition. Unless otherwise stated, we set $J=1$.
The measurement outcomes are described by a marked point process \cite{Fuji20, Yamamoto23, Yamamoto25}, whose trajectory-level dynamics is governed by the Ito-type stochastic equation \cite{Daley14}
\begin{align}
    d\rho_c
    =
    dt\mathcal L \rho_c
    +
    \sum_{j=1}^{L}
    (dN_j-dt\gamma\langle n_j\rangle_c)
    \left(
    \frac{n_j\rho_c n_j}{{\langle n_j\rangle_c}}
    -
    \rho_c
    \right)
\label{eq_stochastic}
\end{align}
with $n_j=c_j^\dagger c_j$.
Here, $dN_j=0,1$ counts a quantum jump at site $j$, which satisfies $dN_j dN_k=\delta_{jk}dN_j$ and $\mathbb{E}[dN_j]=\gamma\langle n_j\rangle_c dt$ with $\langle n_j\rangle_c\equiv\mathrm{Tr}[n_j\rho_c]$, $\mathbb{E}[\cdots]$ denotes the ensemble average over the stochastic process, and the superoperator $\mathcal L$ generates the Lindblad dynamics \cite{Lindblad76}
\begin{align}
    \frac{d\rho}{dt}=\mathcal{L}(\rho)
    =
    -i[H,\rho]
    +
    \gamma\sum_{j=1}^{L}
    \left(
    n_j\rho n_j-\frac{1}{2}\{n_j,\rho\}
    \right).
    \label{eq_lindblad}
\end{align}
In Eqs.~\eqref{eq_stochastic} and \eqref{eq_lindblad}, we take the initial state as the infinite-temperature state, $\rho_{\rm ss}\equiv\mathbb{I}/d_0$. Here, $\mathbb{I}$ is the identity matrix and $d_0$ is the dimension of the Hilbert space in the half-filling sector.

The number of quantum jumps and its fluctuations are fundamental dynamical observables of trajectory ensembles, whose statistics characterize dynamical activity and can reveal dynamical phase transitions~\cite{Garrahan10,Lesanovski13,Landi24}.
Our central observable is the number of quantum jumps in a subsystem~\cite{Yamamoto25}, given by
\begin{align}
    N_{M}(L,t)
    \equiv
    \int_{0}^{t} \sum_{j=1}^M dN_j .
\label{eq_Njump}
\end{align}
The corresponding subsystem noise is defined as the time derivative of the variance of Eq.~\eqref{eq_Njump} over trajectories
\begin{gather}
    D_M(L,t)
    \equiv
    \frac{d}{dt}
    V_M(L,t),
    \label{eq_noise}\\
    V_M(L,t)\equiv \mathbb{E}[N_{M}(L,t)^2]-\mathbb{E}[N_{M}(L,t)]^2.
    \label{eq_variance}
\end{gather}
Here, noise is used in the sense that it quantifies the growth rate of the variance of the accumulated jump number, whose long-time limit corresponds to the zero-frequency noise of the jump current~\cite{Landi24}; note that it is different from the stochastic noise entering fluctuating-hydrodynamic or Langevin equations as uncontrollable effects of microscopic fast degrees of freedom.
We should beware that subsystem noise \eqref{eq_noise} is defined from quantum trajectories, but that Eqs.~\eqref{eq_noise} and \eqref{eq_variance} can be expressed through the ensemble-averaged dynamics \eqref{eq_lindblad} \cite{Landi24}.
Introducing the subsystem superoperator $\mathcal{J}_M\rho\equiv\gamma \sum_{j=1}^M n_j\rho n_j$, the noise is expressed in terms of the Liouvillian autocorrelation function as \cite{Yamamoto25}
\begin{align}
    D_M(L,t)
    =
    K_M
    +
    2\int_0^t d\tau
    \left\{
    {\rm Tr}\left[
    \mathcal{J}_M e^{\mathcal{L}\tau}
    \mathcal{J}_M\rho_{\rm ss}
    \right]
    -
    J_M^2
    \right\}.
    \label{eq_noise2}
\end{align}
Here, $K_M\equiv\gamma\sum_{j=1}^M{\rm Tr}[n_j\rho_{\rm ss}]$ and $J_M\equiv{\rm Tr}[\mathcal{J}_M\rho_{\rm ss}]$, which reduce to $K_M=J_M=\gamma M/2$ at half filling. Thus, the noise naturally decomposes into 
\begin{align}
D_M(L,t)=D_M^{\rm act}(L)+D_M^{\rm anom}(L,t),
\label{eq_decompose}
\end{align} 
with $D_M^{\rm act}(L)\equiv K_M$ and
\begin{align}
    D_M^{\rm anom}(L,t)
    &\equiv
    2\gamma^2
    \int_0^t d\tau\,
    \sum_{j,k=1}^M C_L(j-k,\tau),\label{eq:noise_C}
\end{align}
where $C_L(x,t)\equiv\braket{n'_{x+1}(t)n'_1}_\mathrm{ss}$, $n'_j\equiv n_j-\langle n_j\rangle_\mathrm{ss}$, $\langle\cdots\rangle_\mathrm{ss}\equiv\mathrm{Tr}[\cdots\rho_\mathrm{ss}]$, $n_j(t)\equiv e^{\widetilde{\mathcal{L}}t}n_j$ with the adjoint Liouvillian $\widetilde{\mathcal{L}}$, $x\in \mathbb{Z}/L\mathbb{Z}$, and translational invariance of the spatiotemporal correlation function $\langle n'_j(t)n'_k\rangle_\mathrm{ss}=\braket{n'_{j-k+1}(t)n'_1}_\mathrm{ss}$ is used.
We note that the variance also decomposes into $V_M(L,t)= V_M^{\rm act}(L,t)+V_M^{\rm anom}(L,t)$ corresponding to Eq.~\eqref{eq_decompose}, where $\frac{d}{dt}V_M^\mathrm{anom\:(act)}\equiv D_M^\mathrm{anom\:(act)}$.
The term $D_M^{\mathrm{act}}(L)$ is called the dynamical activity, which gives the Poissonian contribution \cite{Landi24}, while $D_M^{\rm anom}(L,t)$ is the anomalous super-Poissonian contribution originating from the integrated autocorrelation function \cite{Yamamoto25}. Here, we call the fluctuations super-Poissonian when $V_M>\mathbb{E}[N_M]$. We remark that the anomalous term and nontrivial dynamic scaling are absent for the whole system ($M=L$) because the total particle number is a conserved quantity \cite{total}. 
The dynamic scaling discussed below is thus controlled by the anomalous term.

\textit{Exact dynamic scaling}--In the following, we exactly evaluate the noise $D_M^\mathrm{anom}(L,t)$, whose dynamic scaling also indicates the dynamic scaling of $D_M(L,t)$, $V_M^\mathrm{anom}(L,t)$, and $V_M(L,t)$ in the large spacetime scales.
First, we rewrite the initial condition as
$
\rho_\mathrm{ss} = \frac{[(L/2)!]^2}{L!} \sum_{1\leq x_1<\cdots <x_{L/2}\leq L} \ket{\bm{x}}\bra{\bm{x}}.
$
To compute $C_L(x,t)$, we introduce the following matrix element
\begin{align}
F_{x_1,x_2}(t)\equiv
\mathrm{Tr}[c^\dagger_{x_1+1}c_{x_2+1}e^{\mathcal{L}t}(n_1\rho_\mathrm{ss})]-\frac{1}{4}\delta_{x_1,x_2},
\end{align}
where $x_1,x_2 \in \mathbb{Z}/L\mathbb{Z}$ and $C_L(x,t)$ is obtained from the diagonal component as 
\begin{align}
C_L(x,t)=F_{x,x}(t).
\label{eq_correlationfun}
\end{align}
A direct calculation using the Lindblad equation \eqref{eq_lindblad} shows that $F_{x_1,x_2}(t)$
satisfies the closed equation~\cite{Eisler11, Medvedyeva16}
\begin{align}
&i\frac{d}{dt}F_{x_1,x_2}(t)=F_{x_1+1,x_2}(t)+F_{x_1-1,x_2}(t)-F_{x_1,x_2+1}(t)\notag\allowdisplaybreaks\\
&-F_{x_1,x_2-1}(t)+i\gamma(\delta_{x_1,x_2}-1)F_{x_1,x_2}(t),
\label{eq:F_evo}
\end{align}
with the initial condition 
\begin{align}
F_{x_1,x_2}(0)=\delta_{x_1,x_2}(L\delta_{x_1,0}-1)/[4(L-1)].
\label{eq:F_ini}
\end{align}
The initial-value problem can be solved by an exact reduction to momentum sectors (see Supplemental Material~\cite{Suppl}).
We thereby obtain the exact solution of the noise as
\begin{widetext}
\begin{equation}
D^{\mathrm{anom}}_M(L,t) = \frac{\gamma^2}{2(L-1)}
\sum_{m=1}^{L-1} \frac{\sin^2(Mm\pi/L)}{\sin^2(m\pi/L)} \int_0^t d\tau\,e^{-\gamma \tau } \oint_{\Gamma_m}\frac{du}{2\pi i}\,
e^{-i\kappa_m \tau u} \frac{\kappa_m U_{L-1}(u)}{\kappa_m[T_L(u)-\sigma_m]-i\gamma U_{L-1}(u)},
\label{eq:noise_exact}
\end{equation}
\end{widetext}
where $T_n$ and $U_n$ denote the Chebyshev polynomials of the first and second kinds, respectively, 
and $\Gamma_m$ is a positively oriented contour enclosing all poles of the integrand. 
We have also introduced $\kappa_m\equiv4\sin(m\pi/L)$ 
and $\sigma_m\equiv(-1)^{m+L/2}$ for $m=1,\ldots,L-1$.

\begin{figure}[b]
\includegraphics[width=8.5cm]{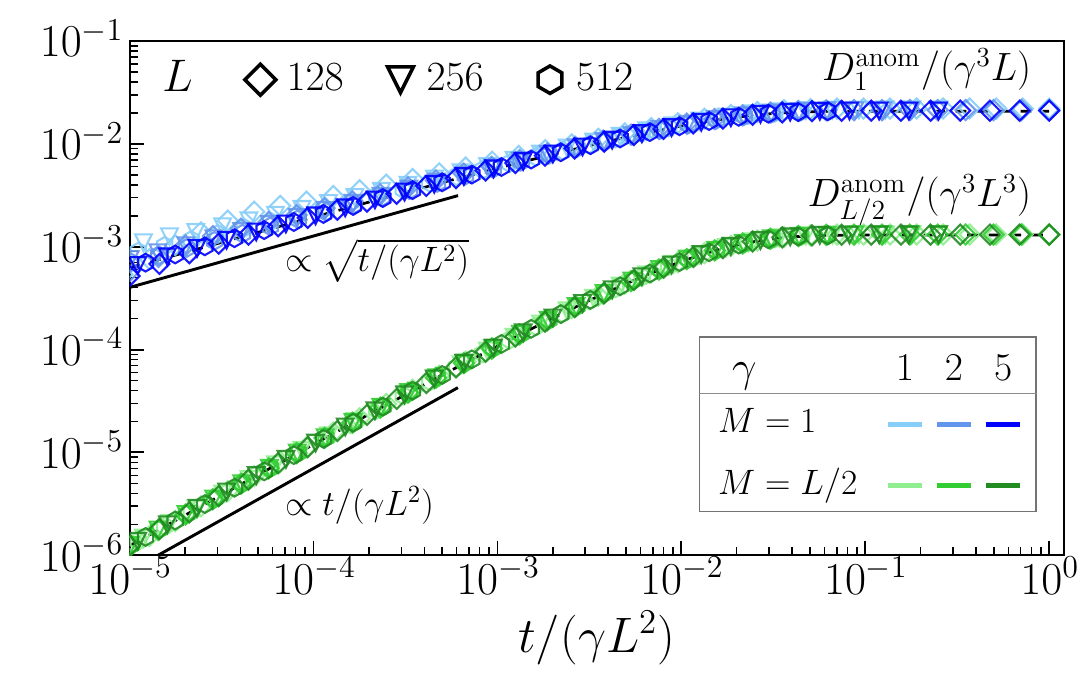}
\caption{Dynamic scaling of $D^{\mathrm{anom}}_M(L,t)$ ($M=L/2,1$), obtained from the exact results \eqref{eq:asym_loc} and \eqref{eq:asym_half} together with the numerical simulation using Eqs.~\eqref{eq_correlationfun}-\eqref{eq:F_ini}. The noise exhibits the dynamic scaling characterized by $D_{M}^\mathrm{anom}(L,t)=\gamma^3 L^\alpha f_M(t/(\gamma L^2))$.}
\label{fig_Dscaling}
\end{figure}

This expression allows us to derive our main result, i.e., the dynamic scaling of $D^{\mathrm{anom}}_M(L,t)$ (for $M=1,L/2$) in the limit $L,t \to \infty$  with fixed $\gamma >0$ and $\xi \equiv t/(\gamma L^2)$ (see Supplemental Material \cite{Suppl} for the calculation):
\begin{align}
&\lim_{L\to\infty} \frac{D^{\mathrm{anom}}_{1}(L,\gamma L^2\xi)}{\gamma ^3 L} = f_1(\xi)\equiv  \sum_{m>0} \frac{1- e^{-8m^2 \pi^2 \xi}}{8m^2\pi^2},
\label{eq:asym_loc}\allowdisplaybreaks\\
&\lim_{L\to\infty} \frac{D^{\mathrm{anom}}_{L/2}(L,\gamma L^2\xi)}{\gamma^3 L^3} = f_{\mathrm{half}}(\xi)\equiv \sum_{m>0} \frac{1 - e^{-8 (2m-1)^2\pi^2 \xi}}{8(2m-1)^4 \pi^4}.
\label{eq:asym_half}
\end{align}
Here, the subscript $\mathrm{half}$ labels the half-chain ($M=L/2$) subsystem.
These scaling functions manifest the following asymptotic behavior
\begin{align}
&f_1(\xi) \simeq 
\begin{cases}
\frac{1}{2} \sqrt{\frac{\xi}{2\pi}}, & \quad(\xi \ll 1),
\\
\frac{1}{48}, & \quad(\xi \gg 1),
\end{cases}
\label{eq:asym_f}
\allowdisplaybreaks\\
&f_{\mathrm{half}}(\xi) \simeq 
\begin{cases}
\frac{\xi}{8}, \qquad &\quad(\xi\ll1),
\\
\frac{1}{768}, &\quad(\xi \gg 1).
\end{cases}
\label{eq:asym_g}
\end{align}
The dynamic scaling of $D^{\mathrm{anom}}_M(L,t)$ obtained from Eqs.~\eqref{eq:asym_loc} and \eqref{eq:asym_half} is depicted in Fig.~\ref{fig_Dscaling} together with the numerical simulation of Eqs.~\eqref{eq_correlationfun}-\eqref{eq:F_ini} using the RK4 method, and then the time integral in Eq.~\eqref{eq:noise_C} is numerically evaluated to obtain the noise.
The asymptotic behavior in Eqs.~\eqref{eq:asym_f} and \eqref{eq:asym_g} for $\xi\ll1$ and $\xi\gg1$, corresponding to short-time and long-time dynamics, is clearly seen in Fig.~\ref{fig_Dscaling}.

Note that $D_M(L,t)$ for $M=1$ and $M=L/2$ also exhibits the same dynamic scaling as in Eqs.~\eqref{eq:asym_loc} and \eqref{eq:asym_half},
since the term $D_M^{\mathrm{act}} = \gamma M/2$ becomes negligible in the $L,t\to\infty$ limit.
Moreover, the variance $V_M(L,t)$ in Eq.~\eqref{eq_variance} also exhibits the dynamic scaling due to its anomalous part $V_M^\mathrm{anom}(L,t)$.
In fact, since $t$-integration of the dynamic scaling of the noise reads $V_M^\mathrm{anom}(L,t)=\gamma^4 L^{\alpha+2}g_M(t/\gamma L^2)$ with $g_M(\xi)\equiv\int_0^\xi d\zeta f_M(\zeta)$, we find that 
\begin{align}
V_M^\mathrm{anom}(L, t)=\gamma^3 L^\alpha t F_M (t/(\gamma L^2))
\end{align}
with $F_M(\xi)\equiv g_M(\xi)/\xi$.
The asymptotic behavior of the scaling is summarized in Table \ref{tab_asymptotic}.

\begin{table}[tb]
	\centering
	\caption{Summary of the asymptotic behavior of the noise and the variance in the short-time ($\xi\ll1$) and long-time ($\xi\gg1$) dynamics.}
	\label{tab_asymptotic}
     \begin{tabular}{ccccc} \hline \hline
    ~~~ ~~~ &
    ~~~$D_{1}^\mathrm{anom}$~~~ & ~~~$D_{L/2}^\mathrm{anom}$~~~ & ~~~$V_{1}^\mathrm{anom}$~~~ & ~~~$V_{L/2}^\mathrm{anom}$~~~\\ \hline
    $\xi\ll1 \quad$ & $\gamma^{5/2}t^{1/2}$ & $\gamma^2 L t$ & $\gamma^{5/2}t^{3/2}$ & $\gamma^2 L t^2$\\
    $\xi\gg1 \quad$ & $\gamma^3 L$ & $\gamma^3 L^3$ & $\gamma^3 L t$ & $\gamma^3 L^3 t$\\ \hline
    \end{tabular}
\end{table}

Importantly, the dynamic-scaling exponent depends on how $M$ scales with $L$, with distinct exponents for $M=\mathcal O(L^0)$ (e.g., a single site) and $M=\mathcal O(L)$ (e.g., a half chain).
Physically, this difference originates from the spatial summation of the correlation function $C_L(j-k,t)$ in Eq.~\eqref{eq:noise_C}: for a half chain, the sums over $j$ and $k$ provide an additional contribution absent in the single-site case. This contribution is of order $L^2$ in the long-time dynamics, while in the short-time dynamics it is of order $L\ell(t)$, where $\ell(t)$ is a characteristic correlation length, expected to scale diffusively as $\ell(t)\sim\sqrt{t/\gamma}$.

We note that the characteristic time scale is only determined through $t/(\gamma L^z)$ with the dynamic exponent $z=2$, showing that the underlying dynamics is diffusive.
We stress that, while $z=2$ is associated with the Liouvillian gap that scales as $\Delta\sim 1/L^2$ \cite{Gwa92a, Gwa92l, Cai13, Znidaric15}, the remaining exponents can vary depending on the monitored subsystem and are not determined from the Liouvillian gap.

\textit{Analytical solution for interacting fermions for sufficiently strong measurement strength}--Next, we analytically argue that the dynamic scaling relations \eqref{eq:asym_loc} and \eqref{eq:asym_half} even hold in a continuously monitored interacting fermion for $\gamma\gg1$. To this end, we study a fermion chain with the nearest-neighbor interaction, given by $H=-\sum_{j=1}^{L}(
c_{j+1}^\dagger c_j+c_j^\dagger c_{j+1})+V_\mathrm{int}\sum_{j=1}^{L} n_j n_{j+1}$. In Eq.~\eqref{eq_lindblad}, the off-diagonal elements of the density matrix in the Fock basis rapidly decay for $\gamma\gg1$.
Therefore, the effective dynamics in the diagonal subspace is obtained by the second-order perturbation theory.
Since the nearest-neighbor interaction is diagonal in the Fock basis, it does not give a leading-order contribution in the diagonal subspace.
The resulting effective dynamics is described by the symmetric simple exclusion process (SEP)~\cite{Cai13}. 

Specifically, we first introduce the probability distribution $P(x_1,\ldots,x_N;t)\equiv\bra{0}c_{x_N}\cdots c_{x_1}\rho(t)c^\dagger_{x_1}\cdots c^\dagger_{x_N}\ket{0}$ for $x_1<\cdots<x_N$ and identify it with the spin state as $\ket{P(t)}\equiv\sum_{x_1<\cdots<x_N}P(x_1,\ldots,x_N;t)S^+_{x_1}\cdots S^+_{x_N}\ket{0}_\mathrm{spin}$,
where we introduce the spin-$1/2$ operators $S_j^\alpha\;(\alpha=+,-,z)$, and $\ket{0}_{\rm spin}$ denotes the all-down-spin state.
For $\gamma\gg 1$ with fixed $V_\mathrm{int}$, the effective equation of motion is given by $\frac{d}{dt}\ket{P(t)}\simeq \frac{2}{\gamma}\mathcal{L}_{\rm SEP}\ket{P(t)}$
with the Markov generator of the SEP with unit hopping rate $\mathcal{L}_{\mathrm{SEP}}\equiv\sum_{j=1}^{L}[S_j^+S_{j+1}^- + S_j^-S_{j+1}^+ + 2S_j^z S_{j+1}^z -\frac{1}{2}]$.
For the SEP, the equation of motion for the correlation function is closed, leading to
\begin{align}
\frac{d}{dt}
\braket{n_j(t)n_k}_\mathrm{ss}
\simeq
\frac{2}{\gamma}
\big[
&\braket{n_{j+1}(t)n_k}_\mathrm{ss}
+
\braket{n_{j-1}(t)n_k}_\mathrm{ss}\notag\\
&-
2\braket{n_j(t)n_k}_\mathrm{ss}
\big],
\label{eq:SEP_corr_eom}
\end{align}
up to the leading order of $\gamma^{-1}$.
Note that the initial condition is given by $\braket{n_j n_k}_\mathrm{ss}=1/4+(L\delta_{j,k}-1)/[4(L-1)]$. Equation \eqref{eq:SEP_corr_eom} is solved, leading to
\begin{align}
D^{\mathrm{anom}}_M(L,t) 
\simeq
\sum_{m=1}^{L-1}
\frac{\gamma^3\sin^2\left(\frac{\pi m M}{L}\right)[1 - e^{\frac{4t}{\gamma} [\cos (\frac{2m\pi}{L})-1]}]}{8(L-1)\sin^2\left(\frac{\pi m}{L}\right)[1- \cos (\frac{2m\pi}{L})]},
\label{eq:noise_strong_measurement}
\end{align}
which gives the dynamic scaling laws \eqref{eq:asym_loc} and \eqref{eq:asym_half} in the limit $L,t/\gamma \to\infty$ with fixed $\xi$ (see Supplemental Material \cite{Suppl} for the derivation).

\textit{Quantum trajectory simulation}--Quantum trajectory method is one of the most powerful numerical frameworks to simulate continuously monitored dynamics relevant to AMO experiments, where the number of trajectories plays a role of experimental trials \cite{Daley14, Yamamoto25, Yamamoto26}. Because the exact solution cannot address the effect of a finite number of trajectories, whether the dynamic scaling can be observed under realistic finite-sampling conditions is only assessed through quantum trajectory simulations. Here, we employ the quantum trajectory method for a free fermion following the Gaussian-state-based algorithms \cite{Alberton21, Minato22, Muller22, Szyniszewski23, Matsubara25, Matsubara26} with 100 trajectories.
We take the initial state to be the half-filled N\'eel state and evolve it until the saturation of entanglement entropy is reached in each trajectory; then, the steady state density matrix is well approximated by $\rho_\mathrm{ss}$ \cite{Suppl}.
In numerical simulations, we count the number of jumps in Eq.~\eqref{eq_Njump} that is directly accessible in experiments.
We therefore compute the variance $V_M$ in Eq.~\eqref{eq_variance} over trajectory realizations generated by Eq.~\eqref{eq_stochastic}, and compare the results with the exact solution, which is obtained by integrating Eqs.~\eqref{eq:asym_loc} and \eqref{eq:asym_half}.

\begin{figure}[tb]
\includegraphics[width=8.5cm]{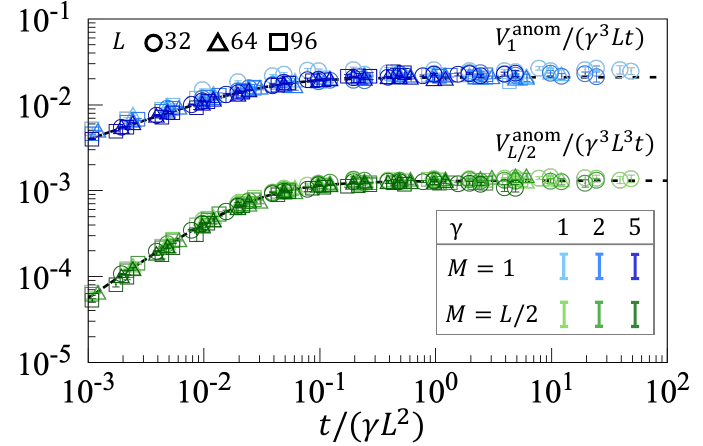}
\caption{Numerical results for the dynamic scaling of $V_{M}^{\mathrm{anom}}(L,t)$ ($M=L/2,1$) based on the quantum trajectory method \cite{dynamic}, and their comparison with the exact results (dashed curve) obtained by integrating Eqs.~\eqref{eq:asym_loc} and \eqref{eq:asym_half}. The variance exhibits the dynamic scaling characterized by $V_{M}^\mathrm{anom}(L,t)=\gamma^3 L^\alpha t F_M(t/(\gamma L^2))$.}
\label{fig_Dynamicscaling}
\end{figure}

In Fig.~\ref{fig_Dynamicscaling}, we show the variance $V_M^\mathrm{anom}$ ($M=1, L/2$) obtained from the quantum trajectory simulation and its comparison with the exact results. Clearly, we find that the numerical results for $V_M^\mathrm{anom}/(\gamma^3 L^\alpha t)$ [$\alpha=1$ ($\alpha=3$) for $M=1$ ($M=L/2$)] exhibit a scaling collapse against $\xi=t/\gamma L^2$, which agrees well with the exact results for both a half chain and a single site. This highlights that $V_M^\mathrm{anom}$ manifests the dynamic scaling discussed in the exact analysis, characterized by $V_M^\mathrm{anom}(L, t)=\gamma^3 L^\alpha t F_M ((t/\gamma L^2))$ \cite{Suppl}.
See Appendix in End Matter for the quantitative comparison of the scaling exponent between numerical and exact results.

Let us discuss how the predicted dynamic scaling can be tested in ultracold-atom experiments. 
We restore the hopping amplitude $J$ here.
For $L\simeq 30$ and $\gamma/J\simeq 1$, e.g., $\xi=\mathcal O( 10^{-1})$ requires an evolution time of $tJ=\mathcal O(10^2)$. For comparison, evolution times up to $tJ\simeq 700$ have been demonstrated in one-dimensional Fermi-Hubbard experiments with $L\simeq 290$ \cite{Scherg21}. 
While continuous monitoring is more experimentally challenging, off-resonant probe light \cite{Luschen17, Patil15} combined with quantum-gas microscopy \cite{Bakr09, Sherson10} provides a possible route toward realizing particle-number measurements, as discussed in Ref.~\cite{Ashida16}. Finally, we again stress that our results are free from the postselection problem.

\textit{Conclusions}--By identifying novel dynamic scaling, we have established the subsystem noise and the variance of quantum jumps as experimentally accessible probes of measurement-induced nonequilibrium universality. For continuously monitored free fermions and interacting fermions for $\gamma\gg1$, we have shown that $D_{M}^{\mathrm{anom}}$ exhibits dynamic scaling laws derived from the exact solution of the Liouvillian autocorrelation function: $D_{M}^\mathrm{anom}(L,t)=\gamma^3 L^\alpha f_M(t/(\gamma L^2))$, where $\alpha$ varies depending on the monitored subsystem. 
Moreover, for free fermions under continuous monitoring, we have elucidated that the variance obeys the dynamic scaling given by $V_M^\mathrm{anom}(L, t)=\gamma^3 L^\alpha t F_M ((t/\gamma L^2))$.

Demonstrating the universality of these exponents in broader classes of monitored many-body models remains an important future direction.
In particular, it would be interesting to examine higher moments and to search for other universality unique to open systems with, e.g.,  nonreciprocity or dissipative phase transitions \cite{Soares26}.
We also mention that Ref.~\cite{Yamamoto25} has obtained the system-size scaling in the long-time dynamics in continuously monitored XXZ models including free fermions, and the extrapolated exponent was $\alpha\simeq 2.7$ up to $L=20$ with $M=L/2$ based on exact diagonalization.
Since our results state that this exponent converges to 3 in the thermodynamic limit in free fermions, it is important to investigate whether the exponent in interacting fermions under finite measurement strengths would approach 3 or not.

\begin{acknowledgments}
\textit{Acknowledgments}--We are grateful to Kenji Shimomura for fruitful discussions. This work was supported by JSPS KAKENHI Grants No.~JP25K17327, No.\ JP25KJ1247, No.\ JP23K13029, No.\ JP23K22414, and No.\ JP24K16982,
Hirose Foundation,
Fujikura Foundation,
Toyota Riken Scholar Program,
Support Center for Advanced Telecommunications Technology Research, Matsuo Foundation, and Mazda Foundation.
K.Y. was supported by JSPS Program for Forming Japan's Peak Research Universities (J-PEAKS) Grant No.~JPJS00420230008.
T.I. and T.M. were supported by JST SPRING Grant No.\ JPMJSP2180.
R.H. was supported by JST ERATO Grant No.\ JPMJER2302.
The simulations were performed in part with the help of the supercomputing system in ISSP, the University of Tokyo.\end{acknowledgments}

\textit{Data availability}--The data that support the findings of this Letter are not publicly available upon publication because it is not technically feasible and/or the cost of preparing, depositing, and hosting the data would be
prohibitive within the terms of this research project. The
data are available from the authors upon reasonable request.

%


\appendix

\renewcommand{\theequation}{A\arabic{equation}}
\renewcommand{\theHequation}{A\arabic{equation}}
\setcounter{equation}{0}

\onecolumngrid

\begin{center}
   \textbf{End Matter}
 \end{center}
 
\twocolumngrid

\begin{figure}[t]
\includegraphics[width=8.5cm]{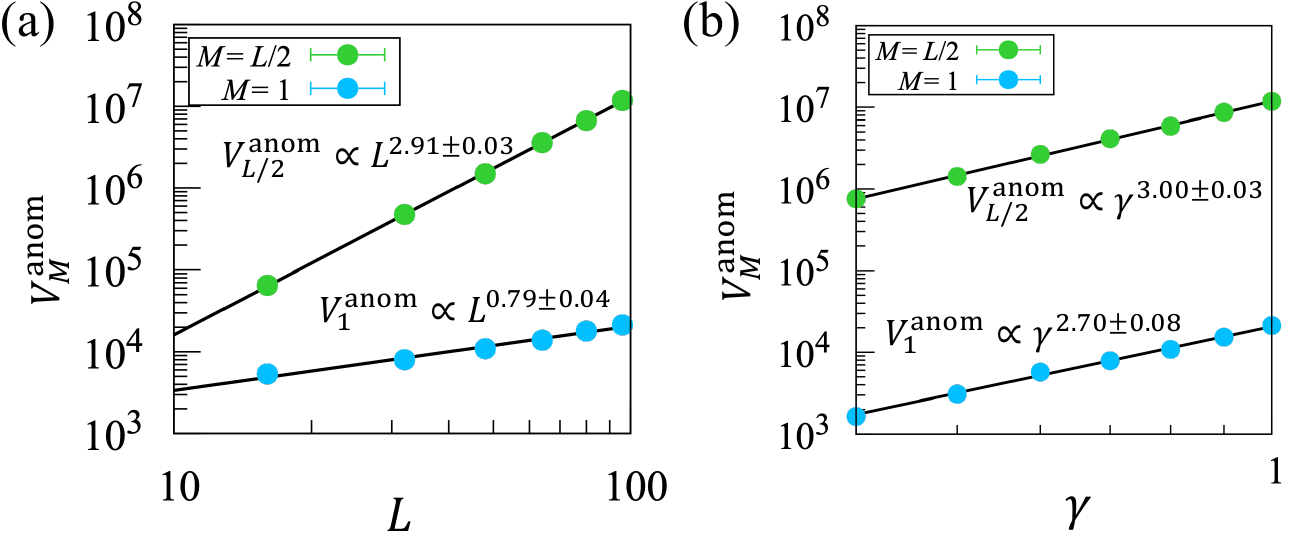}
\caption{Numerical results for $V_{M}^{\mathrm{anom}}$ ($M=L/2,1$) based on the quantum trajectory method for $t=10^4$ (long-time dynamics), depicted against (a) $L$ for $\gamma=1$ and (b) $\gamma$ for $L=96$. We obtain the fit (solid line), $V_1^\mathrm{anom}\propto \gamma^{2.70\pm0.08} L^{0.79\pm0.04}$ and $V_{L/2}^\mathrm{anom}\propto \gamma^{3.00\pm0.03} L^{2.91\pm0.03}$, which are in good agreement with the exact results (see text). We take an average over time intervals by dividing the steady state regime into $20$ parts to obtain the data at $t=10^4$ in panels (a) and (b). Error bars are smaller than the symbol size.}
\label{fig_Tinfinity}
\end{figure}

\textit{Appendix: Asymptotic scaling behavior of $V_M^\mathrm{anom}$ based on the quantum trajectory method}--We analyze the scaling exponents in two limiting regimes $\xi\ll1$ (short-time dynamics) and $\xi\gg1$ (long-time dynamics) and compare the finite-size numerical results based on the quantum trajectory method and exact results given in Table \ref{tab_asymptotic}, showing that the asymptotic scaling behavior is accessible under finite sampling conditions.
In Fig.~\ref{fig_Tinfinity}, we show the numerical results in the long-time dynamics of the scaling behavior of $V_{M}^{\mathrm{anom}}$ ($M=L/2,1$). We set $\xi=t/(\gamma L^2)\simeq1.09$ or larger, noting that the exact solution of the noise almost saturates at a maximum value in this regime (see Fig.~\ref{fig_Dscaling}). From the extracted scaling exponents for the (a) $L$ and (b) $\gamma$ dependences, we find that $V_1^\mathrm{anom}\propto \gamma^{2.70\pm0.08} L^{0.79\pm0.04}$ and $V_{L/2}^\mathrm{anom}\propto \gamma^{3.00\pm0.03} L^{2.91\pm0.03}$. The exact solutions \eqref{eq:asym_f} and \eqref{eq:asym_g} exhibit the scaling behavior $V_{1}^\mathrm{anom}\propto\gamma^3 L$ and $V_{L/2}^\mathrm{anom}\propto\gamma^3 L^3$, where the exponents are in good agreement with the numerical results (note that the trivial $t$ dependence is omitted). Though we find slightly larger deviation for the scaling exponents in $V_1^\mathrm{anom}$, this comes from the fact that number of quantum jumps in the local site is much less than that of the half chain, causing slow numerical convergence.

\begin{figure}[tb]
\includegraphics[width=8.5cm]{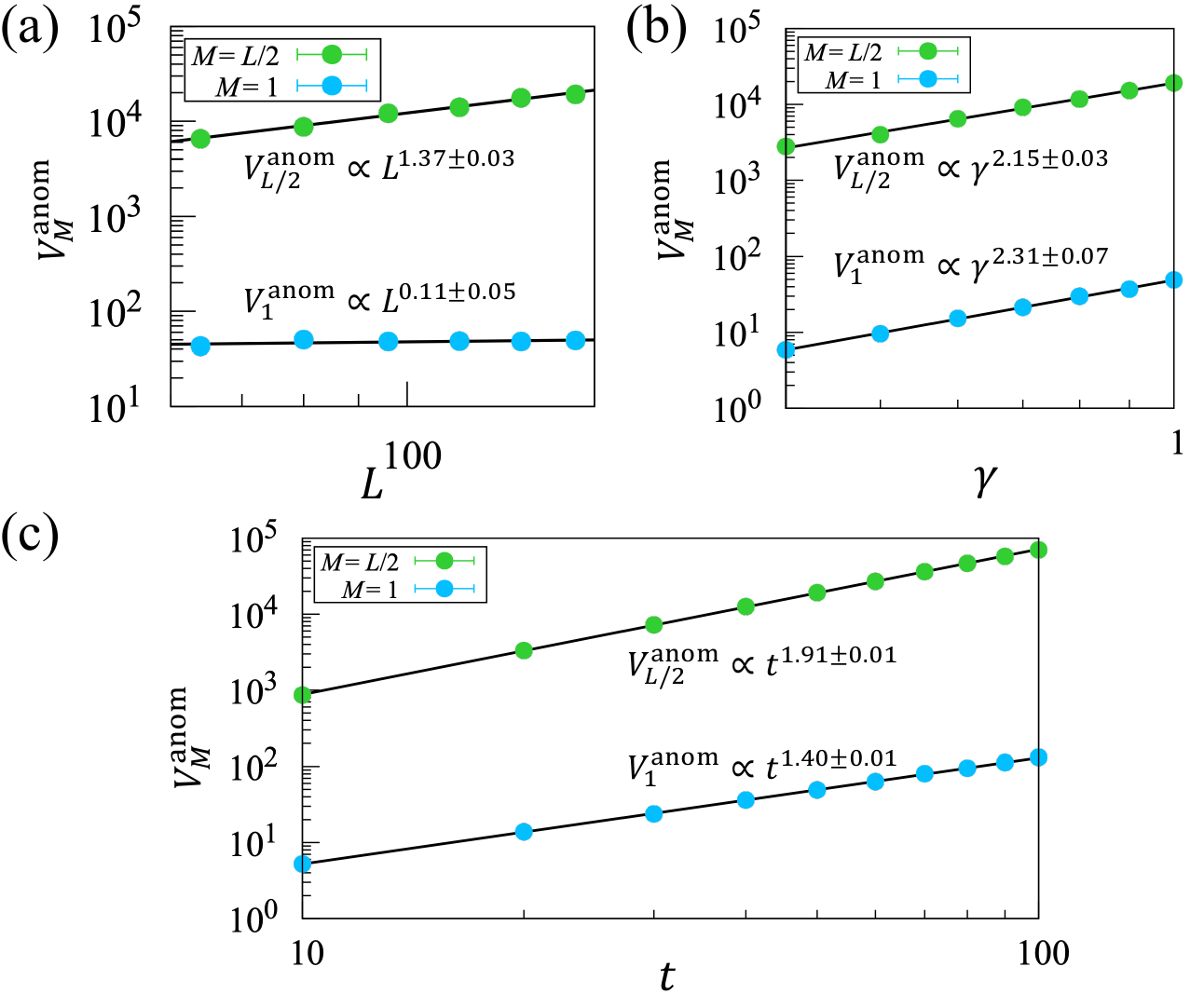}
\caption{Numerical results for $V_{M}^{\mathrm{anom}}$ ($M=L/2,1$) based on the quantum trajectory method in the short-time dynamics. The results are displayed against (a) $L$ for $\gamma=1$ and $t=50$, (b) $\gamma$ for $L=144$ and $t=50$, and (c) $t$ for $\gamma=1$ and $L=144$. We obtain the fit (solid line), $V_1^\mathrm{anom}\propto \gamma^{2.31\pm0.07} L^{0.11\pm0.05} t^{1.40\pm0.01}$ and $V_{L/2}^\mathrm{anom}\propto \gamma^{2.15\pm0.03} L^{1.37\pm0.03} t^{1.91\pm0.01}$, which are in good agreement with the exact results (see text). We take an average over time intervals by dividing the steady state regime into $40$ parts to obtain the data at $t=50$ in panels (a) and (b). In panel (c), we vary the number of such divisions to obtain the $t$ dependence. Error bars are smaller than the symbol size.}
\label{fig_Linfinity}
\end{figure}

We then numerically obtain the scaling form in the short-time dynamics. 
From Figs.~\ref{fig_Linfinity}(a)-(c), we obtain the numerical fit as $V_1^\mathrm{anom}\propto \gamma^{2.31\pm0.07} L^{0.11\pm0.05} t^{1.40\pm0.01}$ and $V_{L/2}^\mathrm{anom}\propto \gamma^{2.15\pm0.03} L^{1.37\pm0.03} t^{1.91\pm0.01}$.
Here, as shown in Fig.~\ref{fig_Linfinity}, $t/(\gamma L^2)$ seems to be sufficiently small, e.g., $t/(\gamma L^2)=0.002$ for $L=144$, $t=50$, and $\gamma=1$.
However, we remark that the exact scaling exponents of the variance deviate from those in Table~\ref{tab_asymptotic}. This is because, unlike in the long-time regime where the scaling functions have essentially saturated, the effective scaling exponents still vary appreciably in the short-time regime over the finite parameter range considered here.
Then, the accurate exponents extracted from Eqs.~\eqref{eq:asym_loc} and \eqref{eq:asym_half} read $V_{1}^\mathrm{anom}\propto\gamma^{2.57} L^{0.17}t^{1.46}$ and $V_{L/2}^\mathrm{anom}\propto\gamma^{2.13} L^{1.35}t^{1.92}$, where the range of parameters used for the fit is the same as those in Fig.~\ref{fig_Linfinity}. 
By comparing the numerical results with these exponents, we find that they are consistent with each other.

\clearpage

\onecolumngrid
\appendix

\renewcommand{\thesection}{S\arabic{section}}
\renewcommand{\theequation}{S\arabic{equation}}
\renewcommand{\theHequation}{S\arabic{equation}}
\setcounter{equation}{0}
\renewcommand{\thefigure}{S\arabic{figure}}
\renewcommand{\theHfigure}{S\arabic{figure}}
\setcounter{figure}{0}

\begin{center}
\large{Supplemental Material for}\\
\textbf{``Exact Dynamic Scaling of Quantum Jump Statistics in Monitored Many-Body Systems"}
\end{center}

\section{Derivation of the exact solution \eqref{eq:noise_exact} in the main text}
We solve the initial-value problem given in Eqs.~\eqref{eq:F_evo} and \eqref{eq:F_ini} in the main text.
To solve this problem, we first exploit translational invariance to decompose the dynamics into independent momentum sectors, each mapped onto an effective single-particle problem.
We then construct the local resolvent of this problem and obtain the relevant return amplitude from its contour-integral representation.
For related exact studies of dynamical observables in dephasing noninteracting fermions, 
see Refs.~\cite{Ishiyama25stat,Bhat25,Ishiyama25fluctuation, Penc26, Essler26}.

We introduce the relative coordinate $r=x_1-x_2$ and the simultaneous-translation coordinate $y=x_2$, and define
\begin{align}
F_m(r,t) &\equiv \sum_{y=0}^{L-1} e^{-iK_my} F_{r+y,y}(t), \qquad K_m=\frac{2\pi m}{L}. \label{eq:Fm_def}
\end{align}
Equation~\eqref{eq:F_evo} in the main text then decomposes into independent momentum sectors as
\begin{align}
i\frac{d}{dt} F_m(r,t)
=(1-e^{-iK_m})F_m(r+1,t) +(1-e^{iK_m})F_m(r-1,t) +i\gamma(\delta_{r,0}-1)F_m(r,t).
\label{eq:Fm_evo}
\end{align}
The initial condition \eqref{eq:F_ini} in the main text becomes
\begin{align}
F_0(r,0)&=0, 
\label{eq:Fm_ini0}\\
F_m(r,0)&=\frac{L}{4(L-1)}\delta_{r,0}, \qquad m=1,\ldots,L-1.
\label{eq:Fm_ini}
\end{align}
For $m=1,\ldots,L-1$, we define
\begin{align}
\kappa_m \equiv 4\sin\frac{K_m}{2} = 4\sin\frac{m\pi}{L},
\label{eq:kappa_m}
\end{align}
and perform the gauge transformation
\begin{align}
F_m(r,t) = e^{i(K_m/2-\pi/2)r}\phi_m(r,t).
\label{eq:gauge_finite}
\end{align}
The equation of motion is then rewritten as
\begin{align}
i\frac{d}{dt} \phi_m(r,t) = \frac{\kappa_m}{2} [\phi_m(r+1,t)+\phi_m(r-1,t)] +i\gamma(\delta_{r,0}-1)\phi_m(r,t).
\label{eq:phi_evo}
\end{align}
This is a one-particle problem generated by
\begin{align}
h_m = \frac{\kappa_m}{2}(\mathsf T_++\mathsf T_-) -i\gamma +i\gamma\ket{0}\bra{0},
\label{eq:h_m}
\end{align}
where $\mathsf T_\pm$ are the shift operators defined by $\mathsf T_\pm\ket{r}=\ket{r\pm1}$.  
Since $F_m(r+L,t)=F_m(r,t)$, the gauge-transformed wave function obeys the twisted boundary condition
\begin{align}
\phi_m(r+L,t) = \sigma_m\phi_m(r,t), \qquad \sigma_m=(-1)^{m+L/2}, 
\label{eq:twisted_bc}
\end{align}
where we have used that $L$ is even.

We next define the return amplitude 
\begin{align}
A_m(t) \equiv \bra{0}e^{-ih_mt}\ket{0}. \label{eq:A_return}
\end{align}
Then, the inverse Fourier transform of Eq.~\eqref{eq:Fm_def}, together with $F_0(r,t)=0$ that follows from Eqs.~\eqref{eq:Fm_evo} and \eqref{eq:Fm_ini0}, yields
\begin{align}
C_L(x,t) = F_{x,x}(t) =
\frac{1}{4(L-1)} \sum_{m=1}^{L-1} e^{iK_mx}A_m(t).
\label{eq:C_A}
\end{align}
Substituting this expression into Eq.~\eqref{eq:noise_C} in the main text and using
\begin{align}
\sum_{j,k=1}^{M}e^{iK_m(j-k)} = \frac{\sin^2(MK_m/2)}{\sin^2(K_m/2)},
\end{align}
we obtain
\begin{align}
D_M^{\mathrm{anom}}(L,t) = \frac{\gamma^2}{2(L-1)} \sum_{m=1}^{L-1} \frac{\sin^2(Mm\pi/L)}{\sin^2(m\pi/L)} \int_0^t d\tau\,A_m(\tau).
\label{eq:D_A}
\end{align}

It remains to evaluate $A_m(t)$.  We decompose $h_m$ as
\begin{align}
h_m=h_{0,m}+W, \qquad h_{0,m} = \frac{\kappa_m}{2}(\mathsf T_++\mathsf T_-)-i\gamma,
\qquad
W=i\gamma\ket{0}\bra{0}.
\end{align}
Let
\begin{align}
g_m(z)\equiv\bra{0}(z-h_{0,m})^{-1}\ket{0},
\qquad
G_m(z)\equiv\bra{0}(z-h_m)^{-1}\ket{0}.
\end{align}
The resolvent identity $(z-h_m)^{-1}=(z-h_{0,m})^{-1}+(z-h_{0,m})^{-1}W(z-h_m)^{-1}$ gives 
\begin{align}
G_m(z)= \frac{g_m(z)}{1-i\gamma g_m(z)}.
\label{eq:Dyson_local}
\end{align}
The free Hamiltonian $h_{0,m}$ is diagonalized using plane-wave states
$\ket{q_a} = \frac{1}{\sqrt L}\sum_{j} e^{i q_a j }\ket{j}$, where the allowed momenta $q_a$ satisfy the twisted boundary condition $e^{iq_aL}=\sigma_m$.
The corresponding eigenvalues are $\epsilon_m(q_a)=\kappa_m\cos q_a-i\gamma$.
Since $|\braket{0|q_a}|^2=1/L$, the local free resolvent is given by
\begin{align}
g_m(z)= \frac{1}{L} \sum_{a=0}^{L-1} \frac{1}{z+i\gamma-\kappa_m\cos q_a}.
\label{eq:g_sum}
\end{align}

We can evaluate the momentum sum in Eq.~\eqref{eq:g_sum} in closed form
using the Chebyshev polynomials.
The Chebyshev polynomials of the first and second kinds, $T_n(u)$ and $U_n(u)$, are defined by
\begin{equation}
T_n(\cos q) = \cos(nq),\qquad U_n(\cos q) = \frac{\sin[(n+1)q]}{\sin q},
\label{eq:def_T_U}
\end{equation}
and satisfy
\begin{equation}
T_n'(u)=nU_{n-1}(u).
\label{eq:Chebyshev_derivative}
\end{equation}
We also use the logarithmic-derivative identity for a polynomial $P_n(u)=C\prod_{j=1}^{n}(u-\lambda_j)$,
given by
\begin{equation}
\frac{P_n'(u)}{P_n(u)} = \sum_{j=1}^{n}\frac{1}{u-\lambda_j}.
\label{eq:log_derivative}
\end{equation}
The definition of $T_n(u)$ and the twisted boundary condition $e^{iq_aL}=\sigma_m$ imply
\begin{equation}
T_L(\cos q_a) = \cos(Lq_a) = \sigma_m.
\end{equation}
Therefore, $\cos q_a$ ($a = 0,\ldots, L-1$) are the roots of the polynomial $T_L(u)-\sigma_m$.
Applying Eq.~\eqref{eq:log_derivative} to this polynomial, we obtain
\begin{equation}
\sum_{a=0}^{L-1}
\frac{1}{u-\cos q_a}
=
\frac{T_L'(u)}{T_L(u)-\sigma_m}
=
\frac{L U_{L-1}(u)}{T_L(u)-\sigma_m}.
\label{eq_polyTU}
\end{equation}
Setting $u = (z+i\gamma)/\kappa_m $ and substituting Eq.~\eqref{eq_polyTU} into Eq.~\eqref{eq:g_sum}, we obtain
\begin{equation}
g_m(z)
=
\frac{1}{\kappa_m}
\frac{U_{L-1}(u)}
{T_L(u)-\sigma_m}.
\label{eq:g_chebyshev}
\end{equation}
Combining Eqs.~\eqref{eq:Dyson_local} and \eqref{eq:g_chebyshev}, we obtain
\begin{align}
G_m(z) = \frac{U_{L-1}(u)}{\kappa_m[T_L(u)-\sigma_m]-i\gamma U_{L-1}(u)}.
\label{eq:G_chebyshev}
\end{align}

Finally, the propagator \eqref{eq:A_return} can be represented by the resolvent as
\begin{align}
A_m(t) = \oint_{\Gamma_z} \frac{dz}{2\pi i} e^{-izt}G_m(z), \label{eq:A_z}
\end{align}
where $\Gamma_z$ is any positively oriented contour enclosing the spectrum of $h_m$ without crossing it.  
With $z=\kappa_m u-i\gamma$, this becomes
\begin{align}
A_m(t) = e^{-\gamma t} \oint_{\Gamma_m}\frac{du}{2\pi i} e^{-i\kappa_mtu} 
\frac{\kappa_mU_{L-1}(u)}{\kappa_m[T_L(u)-\sigma_m]-i\gamma U_{L-1}(u)}.
\label{eq:A_u}
\end{align}
Combining Eq.~\eqref{eq:A_u} with Eq.~\eqref{eq:D_A} gives the exact finite-$L$ solution \eqref{eq:noise_exact} in the main text.

\section{Derivation of Eqs.~\eqref{eq:asym_loc}-\eqref{eq:asym_g} in the main text: asymptotic behavior of the noise}
We first derive Eqs.~\eqref{eq:asym_loc} and \eqref{eq:asym_half} in the main text from the exact finite-$L$ solution \eqref{eq:noise_exact}, using the contour-integral representation \eqref{eq:A_u}. We define
\begin{equation}
R_{L,m}(u)\equiv\frac{T_L(u)-\sigma_m}{U_{L-1}(u)}.
\label{eq:R_Lm}
\end{equation}
For $u\in\mathbb C\setminus[-1,1]$, we introduce
\begin{equation}
\zeta(u)\equiv u+\sqrt{u^2-1},
\end{equation}
where the branch of $\sqrt{u^2-1}$ is chosen such that $\sqrt{u^2-1}\sim u$ as $|u|\to\infty$. From Eq.~\eqref{eq:def_T_U}, it follows that
\begin{equation}
T_L(u)=\frac{\zeta(u)^L+\zeta(u)^{-L}}{2},
\qquad
U_{L-1}(u)=\frac{\zeta(u)^L-\zeta(u)^{-L}}{2\sqrt{u^2-1}}.
\end{equation}
Thus,
\begin{equation}
R_{L,m}(u)=\sqrt{u^2-1}\,
\frac{1-2\sigma_m\zeta(u)^{-L}+\zeta(u)^{-2L}}
{1-\zeta(u)^{-2L}}.
\label{eq:R_L_exact}
\end{equation}
With this choice of branch, $|\zeta(u)|>1$ for $u\in\mathbb C\setminus[-1,1]$. Hence, for fixed $u$,
\begin{equation}
R_{L,m}(u)=\sqrt{u^2-1}\left[1+\mathcal O\!\left(|\zeta(u)|^{-L}\right)\right].
\label{eq:R_L_asym}
\end{equation}

We now consider the scaling limit $L,t\to\infty$ with fixed $\gamma>0$ and $\xi=t/(\gamma L^2)$. 
From Eq.~\eqref{eq:R_L_asym}, the leading large-$L$ form of Eq.~\eqref{eq:A_u} is
\begin{equation}
A_m(t)\simeq e^{-\gamma t}\oint_{\Gamma}\frac{du}{2\pi i}\,
e^{-i\kappa_mtu}\frac{\kappa_m}{\kappa_m\sqrt{u^2-1}-i\gamma},
\label{eq:A_lim_contour}
\end{equation}
where $\Gamma$ encloses the branch cut $[-1,1]$ and the isolated pole of the integrand, which exists for $\gamma/\kappa_m>1$.
By deforming the contour around the branch cut and the isolated pole, we can separate their contributions.
The former is $\mathcal{O}(e^{-\gamma t})$, while the latter is
\begin{equation}
A_m^{\rm pole}(t)=\frac{\gamma}{\sqrt{\gamma^2-\kappa_m^2}}\exp\left[-\left(\gamma-\sqrt{\gamma^2-\kappa_m^2}\right)t\right].
\label{eq:A_pole}
\end{equation}
After the time integration in Eq.~\eqref{eq:noise_exact}, the branch-cut contribution remains subleading in the scaling limit and therefore does not affect Eqs.~\eqref{eq:asym_loc} and \eqref{eq:asym_half} in the main text.
In the scaling limit, the pole contribution remains of order unity only for $m=\mathcal{O}(1)$, or, symmetrically, for $L-m=\mathcal{O}(1)$. For fixed $m$, we obtain
\begin{equation}
\kappa_m=4\sin\frac{m\pi}{L}=\frac{4\pi m}{L}+\mathcal{O}(L^{-3}),
\end{equation}
and hence
\begin{equation}
A_m(t)\simeq\exp\left(-\frac{\kappa_m^2t}{2\gamma}\right)\simeq\exp\left(-\frac{8\pi^2m^2t}{\gamma L^2}\right).
\label{eq:A_pole_asym}
\end{equation}
For $t=\gamma L^2\xi$, we therefore obtain
\begin{equation}
A_m(\gamma L^2\xi)\longrightarrow e^{-8\pi^2m^2\xi}.
\label{eq:A_scaling_limit}
\end{equation}
We note that the above asymptotic analysis assumes fixed $\gamma>0$, and 
for finite $L$, it requires $\gamma L\gg1$.
When $\gamma=\mathcal O(L^{-1})$, the branch-cut contribution is no longer
exponentially suppressed and the expansion in $\kappa_m/\gamma$ for the pole contribution breaks down,
leading to finite-size corrections to the scaling-limit results.

Thus, substituting Eq.~\eqref{eq:A_scaling_limit} into Eq.~\eqref{eq:noise_exact} in the main text, we obtain, for fixed $m$,
\begin{equation}
\int_0^{\gamma L^2\xi}dt\,A_m(t)\simeq\frac{\gamma L^2}{8\pi^2m^2}\left(1-e^{-8\pi^2m^2\xi}\right).
\end{equation}
Taking into account the identical contribution from $L-m=\mathcal{O}(1)$, for $M=1$ we find
\begin{equation}
\lim_{L\to\infty}\frac{D_1^{\rm anom}(L,\gamma L^2\xi)}{\gamma^3L}
= \sum_{m=1}^{\infty}\frac{1-e^{-8\pi^2m^2\xi}}{8\pi^2m^2},
\end{equation}
which gives Eq.~\eqref{eq:asym_loc} in the main text.
For $M=L/2$, the spatial factor satisfies
\begin{equation}
\frac{\sin^2(Mm\pi/L)}{\sin^2(m\pi/L)} =\frac{\sin^2(m\pi/2)}{\sin^2(m\pi/L)},
\end{equation}
so that only odd $m$ contribute to $D_{L/2}^\mathrm{anom}(L,t)$. Using again the contributions from both $m=\mathcal{O}(1)$ and $L-m=\mathcal{O}(1)$, we obtain
\begin{equation}
\lim_{L\to\infty}\frac{D_{L/2}^{\rm anom}(L,\gamma L^2\xi)}{\gamma^3L^3}
= \sum_{m=1}^{\infty}\frac{1-e^{-8(2m-1)^2\pi^2\xi}}{8(2m-1)^4\pi^4},
\end{equation}
which gives Eq.~\eqref{eq:asym_half} in the main text.

We next derive the asymptotic behavior of the scaling functions, $f_1(\xi)$ and $f_{\mathrm{half}}(\xi)$, given in Eqs.~\eqref{eq:asym_f} and \eqref{eq:asym_g} in the main text.
Let us consider the asymptotic behavior of $f_1(\xi)$.
For $\xi \ll 1$, 
we can replace the sum in Eq.~\eqref{eq:asym_loc} in the main text by an integral with the integration variable $s = 2\sqrt{2} m \pi \sqrt{\xi}$ as
\begin{align}
f_1(\xi) \simeq  \frac{\sqrt{\xi}}{2\sqrt{2}\pi } \int_0^\infty ds \frac{1 - e^{-s^2}}{s^2} = \frac{1}{2}\sqrt{\frac{\xi}{2\pi}}.
\end{align}
For $\xi \gg 1$, we have 
\begin{align}
f_1(\xi) \simeq \sum_{m>0} \frac{1}{8m^2 \pi^2} = \frac{1}{48}.
\end{align}
Thus, we obtain the asymptotic behavior of $f_1(\xi)$ in Eq.~\eqref{eq:asym_f} in the main text.

We then consider the asymptotic behavior of $f_{\mathrm{half}}(\xi)$.
For $\xi \ll 1$, a naive replacement of the sum by an integral is not directly applicable.
Indeed, by introducing
\begin{align}
s_m=2\sqrt{2}(2m-1)\pi\sqrt{\xi},
\end{align}
the summand in Eq.~\eqref{eq:asym_half} in the main text is rewritten as
\begin{align}
\frac{1-e^{-8(2m-1)^2\pi^2 \xi}}{8(2m-1)^4\pi^4}
=8\xi^2 \frac{1-e^{-s_m^2}}{s_m^4}.
\end{align}
The function $(1-e^{-s^2})/s^4$ behaves as $1/s^2$ close to $s=0$,
and thereby the integral approximation contains a singular contribution from the lower boundary.
To isolate this contribution, we add and subtract the singular part as
\begin{align}
f_{\mathrm{half}}(\xi)=\sum_{m>0} \frac{\xi}{\pi^2(2m-1)^2}
-8\xi^2\sum_{m>0}\frac{s_m^2-(1-e^{-s_m^2})}{s_m^4}.
\end{align}
The first term gives the boundary contribution
\begin{align}
\sum_{m>0} \frac{\xi}{\pi^2(2m-1)^2}
=
\frac{\xi}{\pi^2}\sum_{m>0}\frac{1}{(2m-1)^2}
=
\frac{\xi}{8}.
\end{align}
On the other hand, the second term is regular at $s=0$, since
\begin{align}
s^2-(1-e^{-s^2})=\mathcal O (s^4).
\end{align}
Thus, this term can be estimated by an integral and is of order $\xi^{3/2}$.
Therefore, the leading asymptotic behavior is evaluated as
\begin{align}
f_{\mathrm{half}}(\xi) \simeq \frac{\xi}{8},
\qquad
( \xi \ll 1).
\end{align}
On the other hand, for $\xi \gg 1$, the exponential corrections are negligible.
Thus, $f_{\mathrm{half}}(\xi)$ reduces to the convergent series
\begin{align}
f_{\mathrm{half}}(\xi) \simeq \frac{1}{8\pi^4}\sum_{m>0}\frac{1}{(2m-1)^4}
= \frac{1}{8\pi^4}\cdot \frac{\pi^4}{96}
= \frac{1}{768},
\end{align}
where we have used
\begin{align}
\sum_{m>0}\frac{1}{(2m-1)^4}=\frac{\pi^4}{96}.
\end{align}
Therefore, the scaling function $f_{\mathrm{half}}(\xi)$ exhibits the asymptotic behavior given in Eq.~\eqref{eq:asym_g} in the main text.

\section{Derivation of Eq.~\eqref{eq:noise_strong_measurement} in the main text and its scaling behavior}
Let $\rho_x(t)$ be the solution of the discrete diffusion equation on the ring $\mathbb{Z}/L\mathbb{Z}$:
\begin{align}
\frac{d}{dt}\rho_x(t)
=
\rho_{x+1}(t)+\rho_{x-1}(t)-2\rho_x(t),
\quad
\rho_x(0)=\delta_{x,0}.
\end{align}
Then, we obtain
\begin{align}
\rho_x(t)
=
\frac{1}{L}
\sum_{m=0}^{L-1}
\exp\left[
2t\left(\cos\frac{2\pi m}{L}-1\right)
\right]
e^{i\frac{2\pi m}{L}x}.
\end{align}
Thus, using Eq.~\eqref{eq:SEP_corr_eom} in the main text, the connected correlation function is given in the limit of large measurement strength $\gamma\gg1$ as
\begin{align}
&\braket{n'_j(t)n'_k}_\mathrm{ss}
\simeq
\frac{1}{4(L-1)}\left[L\rho_{j-k}\left(\frac{2t}{\gamma}\right)-1\right]
=
\frac{1}{4(L-1)}
\sum_{m=1}^{L-1}
e^{
\frac{4t}{\gamma}
\left(\cos\frac{2\pi m}{L}-1\right)
}
e^{i\frac{2\pi m}{L}(j-k)}.
\label{eq:SEP_corr_fourier}
\end{align}
Substituting Eq.~\eqref{eq:SEP_corr_fourier} into Eq.~\eqref{eq:noise_C} in the main text,
we arrive at
\begin{align}
D^{\mathrm{anom}}_M(L,t)
\simeq \frac{\gamma^2}{2(L-1)}
\sum_{m=1}^{L-1}
\sum_{j,k=1}^{M}
e^{i\frac{2\pi m}{L}(j-k)}
\int_0^t dv\;
e^{
\frac{4v}{\gamma}
\left(\cos\frac{2\pi m}{L}-1\right)
}.
\end{align}
Performing the summation and the integration leads to Eq.~\eqref{eq:noise_strong_measurement} in the main text.

Next, we derive the scaling functions of the noise for the strongly measured interacting fermions described by Eq.~\eqref{eq:noise_strong_measurement} in the main text. 
Specifically, we consider the scaling limit:
$
L,t /\gamma\to\infty
$
with fixed $\xi=t/(\gamma L^2)$.
The term
\begin{equation}
\exp\left\{
\frac{4t}{\gamma}
\left[\cos\left(\frac{2\pi m}{L}\right)-1\right]
\right\},
\end{equation}
appearing in Eq.~\eqref{eq:noise_strong_measurement} in the main text is exponentially
small in the scaling limit, except near the boundaries of the summation
range, where $m=O(1)$ or $L-m=O(1)$. Therefore, only these boundary
regions contribute to the sum in the scaling limit. Expanding the
summand in these regions, we obtain Eqs.~\eqref{eq:asym_loc}
and~\eqref{eq:asym_half} in the main text.

\section{Notes on the quantum trajectory simulation}
Here, we give a remark on the quantum trajectory simulation when calculating the variance.
Our protocol corresponds to averaging the variance over the ensemble of pure-state stochastic trajectories, starting from the N\'eel state $|\psi(t_\mathrm{ini}<0)\rangle=|1010\cdots\rangle$ and evolving it until the time that reaches the steady state, which is characterized by the saturation of the entanglement entropy in each trajectory.
We can regard such a steady-state time to be $t=0$.
Then, if we take the derivative of the variance, it seems that the noise in Eq.~\eqref{eq_noise2} in the main text is obtained.
However, precisely speaking, this is different from the analytical formula \eqref{eq_noise2} in the main text by the factor
\begin{align}
\Delta D_M(L,t) =2\int_0^td\tau J_M^2 
- 2\gamma^2\int_0^t d\tau \sum_{i,j\in M}\mathbb E[\langle\psi(\tau) |n_i|\psi(\tau)\rangle \langle \psi(0)| n_j|\psi(0)\rangle],
\end{align}
where $J_M=\mathrm{Tr}[\mathcal J_M \rho_{\mathrm{ss}}]=\gamma M / 2$ and $\mathbb E[\cdots]$ denotes the ensemble average over the stochastic process during times in $ [t_\mathrm{ini},0]$.
By definition, we can approximate $\mathbb E[\ket{\psi(0)}\bra{\psi(0)}]$ as $\rho_\mathrm{ss}$.
Nevertheless, if we assume that $|\psi(0)\rangle$ were taken from the Haar measure, $\Delta D_M(L,t)$ is exponentially suppressed.
Although the corresponding ensemble does not exactly correspond to the Haar measure, we expect that $|\psi(0)\rangle$ is sufficiently randomly distributed, so we have assumed that $\Delta D_M(L,t)$ is small.
Since the numerical results for the dynamic scaling show good agreement with the exact results (see Fig.~\ref{fig_Dynamicscaling} in the main text), we anticipate that $\Delta D_M(L,t)$ is indeed small so that it does not affect the scaling behavior of the variance in quantum trajectory simulations.

\section{Additional results for the dynamic scaling of the variance}
\begin{figure*}[tb]
\includegraphics[width=17cm]{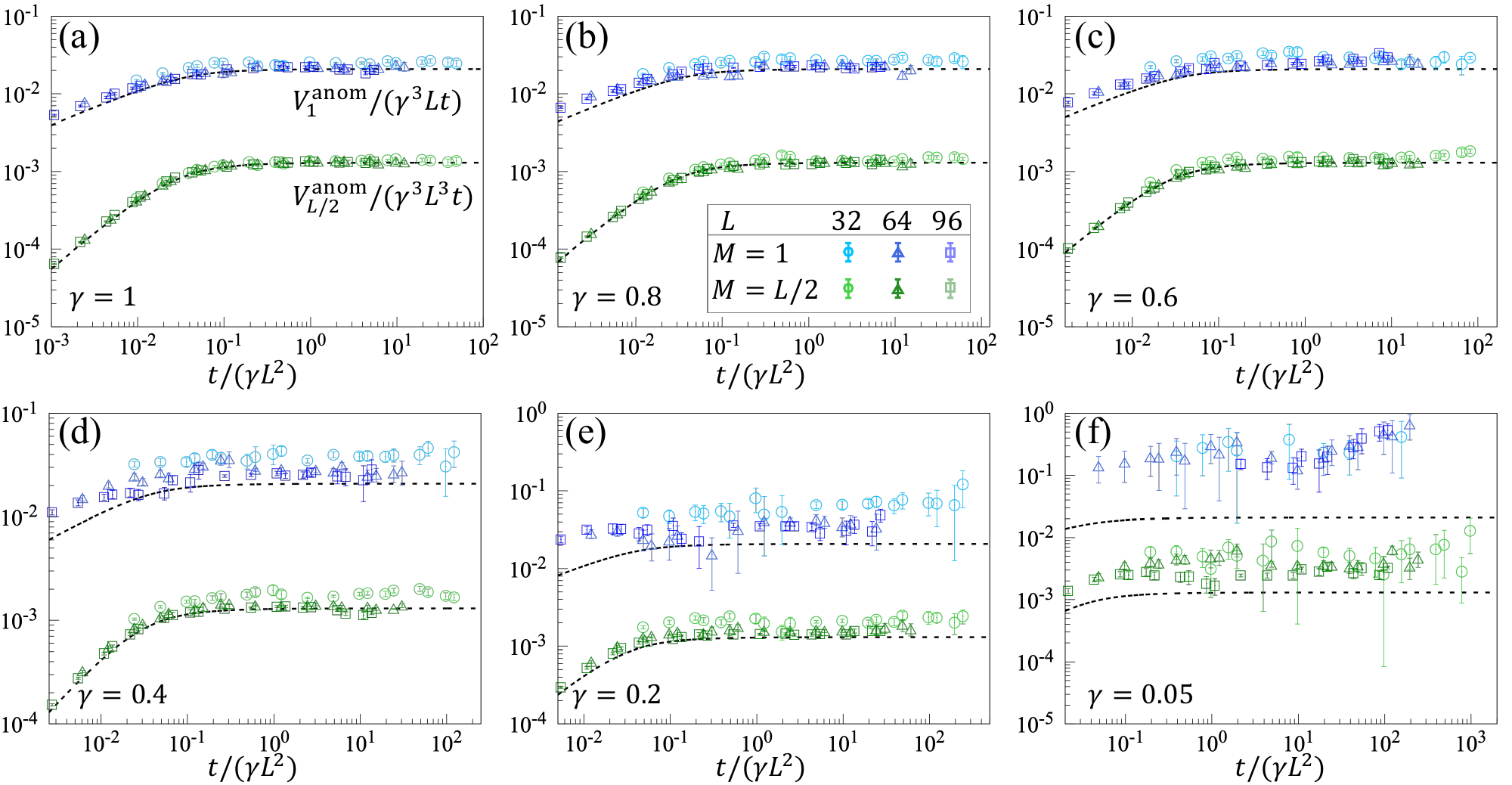}
\caption{Numerical results for the dynamic scaling of $V_{M}^{\mathrm{anom}}$ ($M=L/2,1$) based on the quantum trajectory method, and their comparison with the exact results (dashed curve) obtained by integrating Eqs.~\eqref{eq:asym_loc} and \eqref{eq:asym_half} in the main text. We find that the numerical results agree well with the exact results for $\gamma\sim\mathcal O(1)$, but start to deviate as the measurement strength is decreased, which reflects the validity of application of scaling-limit expressions for the finite-size simulations (see text).}
\label{fig_DynamicscalingAdditional}
\end{figure*}
Figure ~\ref{fig_DynamicscalingAdditional} shows the variance $V_M^\mathrm{anom}$ ($M=1, L/2$) obtained from the numerical results based on the quantum trajectory method, by varying the measurement strength $\gamma$. By changing $t/(\gamma L^2)$, we plot the numerical results together with the exact results, the latter of which is calculated by integrating the dynamic scaling relations \eqref{eq:asym_loc} and \eqref{eq:asym_half} in the main text. In Fig.~\ref{fig_DynamicscalingAdditional}(a), we compare the results for $\gamma=1$, where we find a good agreement for both a half chain and a single site. This demonstrates that $V_M^\mathrm{anom}$ exhibits the dynamic scaling characterized by $V_M^\mathrm{anom}(L, t)=\gamma^3 L^\alpha t F_M ((t/\gamma L^2))$, where $\alpha=1$ for $M=1$ and $\alpha=3$ for $M=L/2$. However, we find that the numerical results start to deviate from the exact results as the measurement strength is decreased [Figs.~\ref{fig_DynamicscalingAdditional}(b)-(f)].
This is because the approach to the scaling limit requires $\gamma L\gg1$
for finite $L$, and thus the weak-measurement regime considered here is
outside the range where the scaling-limit expressions accurately describe
the finite-size results
[see the sentence just after Eq.~\eqref{eq:A_scaling_limit}].
Thus, such a deviation of the numerical simulation is not inconsistent with the exact results.


\begin{thebibliography}{145}
\bibitem [{tot()}]{total}%
  \BibitemOpen
  \href@noop {} {}\bibinfo {note} {Indeed, for the whole system, we obtain
  $D_L(L,t)=\gamma L / 2$ \cite{Yamamoto25}.}\BibitemShut {Stop}%
\bibitem [{dyn()}]{dynamic}%
  \BibitemOpen
  \href@noop {} {}\bibinfo {note} {In quantum trajectory simulations, we take
  an average over time intervals by dividing the steady state regime into
  several parts to obtain sufficient convergence, and error bars calculated
  from averaging over the time window are shown \cite{Yamamoto25}. In all the
  simulations, we set the steady-state time such that entanglement entropy
  exhibits saturation.}\BibitemShut {Stop}%
\end{thebibliography}
\end{document}